# Fusion of Various Optimization Based Feature Smoothing Methods for Wearable and Non-invasive Blood Glucose Estimation

Yiting Wei, Bingo Wing-Kuen Ling*, Danni Chen, Yuheng Dai and Qing Liu
Faculty of Information Engineering, Guangdong University of Technology, Guangzhou 510006, China.

**Abstract:** The traditional blood glucose estimation method requires to take the invasive measurements several times a day. Therefore, it has a high infection risk and the users are suffered from the pain. Moreover, the long term consumable cost is high. Recently, the wearable and non-invasive blood glucose estimation approach has been proposed. However, due to the unreliability of the acquisition device, the presence of the noise and the variations of the acquisition environments, the obtained features and the reference blood glucose values are highly unreliable. Moreover, different subjects have different responses of the infrared light to the blood glucose. To address this issue, this paper proposes a polynomial fitting approach to smooth the obtained features or the reference blood glucose values. In particular, the design of the coefficients in the polynomial is formulated as the various optimization problems. First, the blood glucose values are estimated based on the individual optimization approaches. Second, the absolute difference values between the estimated blood glucose values and the actual blood glucose values based on each optimization approach are computed. Third, these absolute difference values for each optimization approach are sorted in the ascending order. Fourth, for each sorted blood glucose value, the optimization method corresponding to the minimum absolute difference value is selected. Fifth, the accumulate probability of each selected optimization method is computed. If the accumulate probability of any selected optimization method at a point is greater than a threshold value, then the accumulate probabilities of these three selected optimization methods at that point are reset to zero. A range of the sorted blood glucose values are defined as that with the corresponding boundaries points being the previous reset point and this reset point. Hence, after performing the above procedures for all the sorted reference blood glucose values in the validation set, the regions of the sorted reference blood glucose values and the corresponding optimization methods in these regions are determined. It is worth noting that the conventional lowpass denoising method was performed in the signal domain (either in the time domain or in the frequency domain), while our proposed method is performed in the feature space or the reference blood glucose space. Hence, our proposed method can further improve the reliability of the obtained feature values or the reference blood glucose values so as to improve the accuracy of the blood glucose estimation. Moreover, this paper employs the individual modelling regression method to suppress the effects of different users having different responses of the infrared light to the blood glucose values. The computer numerical simulation results show that our proposed method yields the mean absolute relative deviation (MARD) at 0.0930 and the percentage of the test data falling in the zone A of the Clarke error grid at 94.1176%.

## 1. Introduction

The diabetes mellitus is a type of the metabolic diseases characterized by the hyperglycemia. It is a common disease that severely affected the human life and the public health. According to the report issued by the International Diabetes Federation (IDF), an extra of 73.6 millions of people in the worldwide will be suffered from the diabetes at 2045 [1], [26]. As the patients with the diabetes require to take the blood glucose lowering drugs or to inject the insulin to the body for avoiding the occurrence of the diabetes complications, monitoring the blood glucose values is of the great importance to the diabetes.

To monitor the blood glucose values, the traditional method is to have the measurements via the finger prickle approach several times a day. However, this approach has the infection risk [2] and the patients are suffered from the pain. Also, the long term consumable cost is very high. Therefore, there is an urgent need for developing the wearable and non-invasive blood glucose estimation methods [8]. To perform the wearable and non-invasive blood glucose estimation, the optical method is the commonest one among all the methods. This is because this approach allows the continuous data acquisition. It is worth noting that this approach is based on the absorption of the photons by the glucose molecules [3]. Among the whole electromagnetic spectrum, the near infrared spectroscopy shows a good correlation between the blood glucose values and the parameters in the measured photonic response [4], [9], [10]. However, as different subjects have different skin colors, different thicknesses of the fat tissue under the skin, different locations of the blood vessels, different human genes responsible for regulating the blood glucose and different dietary habits, different subjects have different responses of the infrared light to the blood glucose values. Also, since the glucose concentration in the blood is very low, the extracted features are not very sensitive to the blood glucose values. Moreover, there are many substances in the blood vessel. The absorptions of the infrared light by these substances would cause the interferences to the obtained features [5]-[7]. Furthermore, there are the variations in the contact pressure and the contact position exerted on the sensor during the data acquisition. In addition, there is an uncertainty in the acquisition device such as the variations of the component values in the device. Besides, the noise is contaminated to the acquired signals. Apart from that, there are some variations in the acquisition environment such as the variation of the environmental light intensity. From here, it can be concluded that there are two major factors affecting the estimation accuracy. The first one is the variations of the subjects, the devices, the environments and the acquisition conditions. The second one is the interferences from other substances in the blood and the noises. These two types of factors result to the highly unreliable obtained features and the measured reference blood glucose values [11]. Nevertheless, there is no well recognized strategy for addressing these issues in this field.

In many practical data acquisition applications, the regression based smoothing techniques including the linear regression method, the polynomial regression method, the radial basis function regression method and the Fourier basis regression method are used for performing the data pre-processing to eliminate the acquisition errors. Since the order of the polynomial can be changed easily to meet the required specification on the accuracy, this method has a great flexibility. Therefore, the polynomial regression method was widely used in the image processing community, the data processing community and the parameter estimation community. Besides, some parameter design methods were proposed recently. For examples, the full vector based finite element method with an anisotropic perfect matched layer added to the designed model as a boundary condition was employed [28]-[30]. Also, the shooting method (SM) for finding the numerical solutions of the boundary value problems was employed [31]. Moreover, a sign based proportionate affine projection algorithm and a two blocked sparse memory based proportionate affine projection algorithm [33] were used for performing a fast recursive filtering used in the block sparse identification. Furthermore, the dichotomous coordinate descent iteration based method [32] was proposed.

On the other hand, the $L_1$ norm of a vector is the sum of the absolute values of its elements. By minimizing the $L_1$ norm of an error vector, the majorities of the error points are with the small values. As the error vector is sparse, it yields a good estimation for the majority of the data points. Besides, the $L_\infty$ norm of a vector is the maximum absolute value of its elements. By minimizing the $L_\infty$ norm of an error vector, the data points with the large errors are reduced. Recently, the sparse optimization approach and the risk optimization approach were used for performing the data smoothing [12]. Apart from that, the $L_2$ norm of a vector is the square root of the sum of the squares of its elements. The $L_2$ norm error values are between that the $L_1$ norm error values and the $L_\infty$ norm error values. Moreover, as the $L_2$ norm error function is differentiable, the conventional gradient descent approach can be used to find the solution of the optimization problem. Because of this reason, the $L_2$ norm error function is widely used for finding the coefficients of the polynomial for performing the smoothing operation. This is also known as the least squares approach. By applying different approaches with different error norm functions for performing the smoothing operation, different smoothed features are yielded. However, no individual approach is appropriate for estimating all the data points. Hence, the fusion of these three approaches is required. Nevertheless, there is no simple rule for performing the fusion. Hence, this paper is to address this issue.

This paper is to address the unreliability of both the extracted features and the reference blood glucose values. First, this paper employs the individual modelling approach to train the regression system. This can eliminate the effects of different subjects having different responses of the infrared light to the blood glucose values. Second, this paper proposes a method to fuse the various optimization approaches together for performing the polynomial fitting to smooth the features. This can improve the reliability of the extracted features and the reference blood glucose values. Compared to the commonly used data smoothing methods such as the moving averaging method, the Gaussian filtering method, the median filtering method, the locally weighted regression method and the Savitzky Golay (SG) filtering method, the computer numerical simulation results show that our proposed method outperforms the existing methods.

The main contributions of this paper are as follows:
• A feature domain based smoothing method is proposed to improve the reliability of the obtained features and the reference blood glucose values.
• A method is proposed for fusing the various optimization approaches for performing the polynomial fitting to smooth the obtained features and the reference blood glucose values.
• An individual data modelling approach is proposed for performing the blood glucose estimation. This can effectively eliminate the effects of different individuals having different responses of the infrared light to the blood glucose values so as to improve the reliability of the obtained features and the reference blood glucose values.

The outline of this paper is as follows. Section 2 presents our proposed method. Section 3 presents the computer numerical simulation results. Finally, the conclusion is drawn in Section 4.

## 2. Our proposed method
### 2.1. Working principle

The near infrared spectroscopy is a study of the electromagnetic wave. When a beam of photons is hit on the substances in the human tissue [9], [10], the photons are partially absorbed and refracted as well as partially reflected and scattered. Then, the received optical signal is converted to an electrical signal and it is amplified. The electrical signal is called the photoplethysmogram (PPG). It is worth noting that there is a linear correlation between the concentration of a particular substance in the human tissue and the parameters in the near infrared signal. These parameters are related to the spectral absorption. Therefore, the concentrations of the substances in the human tissue can be estimated by extracting the near infrared spectral absorption coefficients from the PPGs [16]. In fact, the non-invasive blood glucose estimation is a typical example of applying the near infrared spectroscopy to the human physiology [4]. However, the human tissue consists of the various substances. In fact, different substances have different ratios of the total number of photons being absorbed to that being reflected for a given beam of photons with a particular wavelength hitting on the human tissue. In particular, for the glucose molecule, the maximum ratio of the total number of photons being absorbed to that being reflected is occurred at the beam of photons with a wavelength being equal to 1600nm [15]. Since the volumes of the skin, the muscle and the bone do not change during the entire circulation cycle of the heartbeat, the absorptions of the optical light due to these substances are constant. On the other hand, the volume of the blood flowing in the artery increases and decreases quasi-periodically during the entire circulation cycle of the heartbeat. Hence, the reflected near infrared light received by the photoelectric receiver also exhibits a quasi-periodical pattern. As a result, the obtained PPG can be represented as the sum of the direct current (DC) component and the alternating current (AC) component. By extracting the AC component out, the characteristics of the blood flow can be characterized. Figure 1 shows the DC component and the AC component of the PPG due to the absorptions of the near infrared light by all the substances. To further investigate this phenomenon, please refer to the Lambert Beer law.

It is worth noting that the autonomic nerves play a vital role in regulating the various physiological activities of the body [17]. However, different diabetic patients have different degrees of the damage in the cardiac autonomic nerves. Hence, the patients with the hyperglycemia have the various degrees of the arrhythmia [18]. To quantify the arrhythmia, the heart rate variability (HRV) is employed. It refers to the variations of the heart rates (HRs). Therefore, the variations of the HRs due to the loss of the regulations of the various physiological activities because of the damage in the cardiac autonomic nerves [19], [21] can be used to estimate the blood glucose levels. To compute the HRV, the conventional algorithms for detecting the QRS points were employed for locating the R waves in the acquired electrocardiograms (ECGs) [22]. Then, the primary method based on generating the velocity maps of the HRV was employed to compute the HRV. Finally, the features derived from the obtained HRV were used to estimate the arterial blood glucose levels [20]. However, as the device for acquiring the ECGs is required to place away from the heart, the blood flows at these positions are relatively slow. Also, the acquisition device has to touch the skin tightly to acquire the ECGs. Nevertheless, the loss of contact is usually happened and it results to the unreliable acquired ECGs. Moreover, the interference will degrade the quality of the acquired ECGs. Hence, it is challenging to estimate the blood glucose level via the acquired ECGs from a practice viewpoint. To address the above difficulty, a more practical method is required for computing the HRV. In particular, the PPG is employed. Figure 2 shows how the PPG is related to the blood flow in the finger. A large number of experimental data shows that the HR estimated by the PPGs is consistent with that estimated by the ECGs when the subjects are in the relaxed state [13]. That is, the interval between the two consecutive peaks (PP interval) estimated by the PPGs is very closed to the RR interval estimated by the ECGs. Moreover, the nonlinear dynamics of the PPGs are very closed to that of the ECGs. Hence, the HRV can be derived using the PPGs. As the PPGs can be acquired more easily [23], [24], [27], this paper only employs the PPGs to perform non-invasive blood glucose estimation.

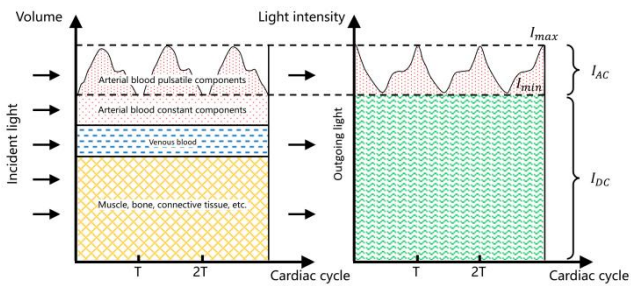

**Figure 1.** *The DC component and the AC component of the PPG due to the different absorptions of the near infrared light by the different substances.*

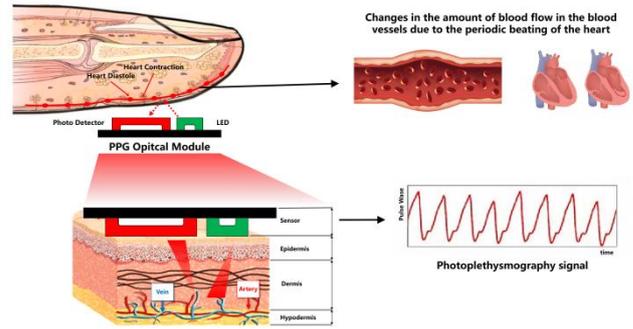

**Figure 2.** *A diagram showing how the PPG is related to the blood flow in the finger.*

Besides using the absorption of the near infrared light and the HRV to estimate the blood glucose levels, the transit time of the PPG was also used to estimate the blood glucose values [14]. In particular, the machine learning approach is used to model the relationship between the transit time of the PPG and the blood glucose values. However, finding the transit time of the PPG is challenging if the synchronous ECG is absence.

2.2. Dataset

Eight volunteers are recruited for performing the data acquisition. In particular, the subjects take the ketogenic diet during the first four days, the normal diet between the fifth day and the eighth day as well as drinking cola after lunch and dinner during the last four days. Figure 3 shows the arrangement of taking these three different types of diets. It is worth noting that taking these three different types of diets will result to the very different blood glucose levels.

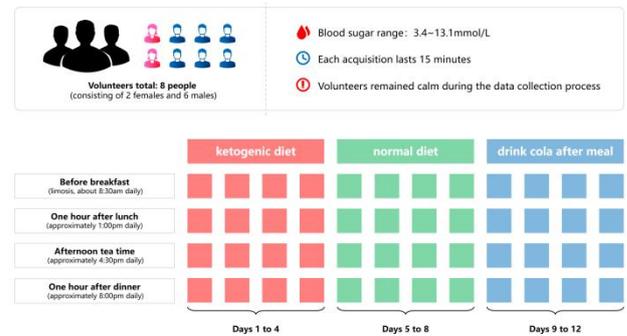

**Figure 3.** *The arrangement of taking three different types of diets.*

2.3. Our proposed algorithm

The block diagram of our proposed algorithm is shown in Figure 4. First, the PPGs are acquired. Second, the acquired PPGs are divided into the training set and the test set. Third, all the PPGs in both the training set and the test set are denoised. Fourth, the features are extracted from each denoised PPG in both the training set and the test set. Fifth, the extracted features or the reference blood glucose values in the training set are smoothed based on the method proposed in this paper. Sixth, some smoothed features are selected. Then, the new lower dimensional feature vectors in both the training set and the test set are formed. Finally, the regression models are built using the new feature vectors and the reference blood glucose values in the training set. Here, the individual modelling approach is employed.

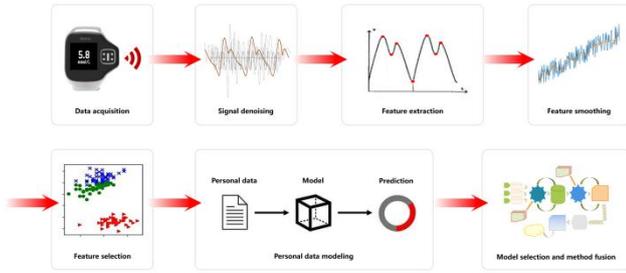

*Figure 4. The block diagram of our proposed method.*

2.3.1. Acquiring the PPGs as well as taking the reference blood glucose values and the reference blood pressure values

Since the heart rates of different subjects are different, the frequency bands of different PPGs acquired from different subjects are different. Hence, an adaptive based denoising method is proposed to address this problem. Among them, the singular spectrum analysis (SSA) based denoising method is the most common adaptive denoising method. However, the conventional SSA based denoising method is to discard the whole SSA component. In this case, some signal information is lost. To address this issue, an SSA based bit plane method is employed for denoising the PPGs. First, the SSA is applied to all the PPGs in both the training set and the test set. Here, the SSA decomposes each PPG into 32 components. Second, for each value in each SSA component, it is represented using a finite number of bits. In this paper, the wordlength of each value is 13 bits. Third, the noise level of each SSA component is estimated. Fourth, the total number of bits corresponding to the noise contaminated to each SSA component is set as the logarithm (based two) of its estimated noise level. Fifth, these noise bits are set to zero. Sixth, each value in each SSA component is reconstructed using the retained bits. Seventh, each PPG is reconstructed by summing up all the processed SSA components together. Eighth, the SSA is performed again on each reconstructed PPG. Ninth, only the first SSA component is retained and it is taken as the denoised PPG. Figure 7 shows a realization of the raw PPG and the corresponding denoised PPG. It can be seen from Figure 7 that the noise is significantly suppressed after performing the denoising.

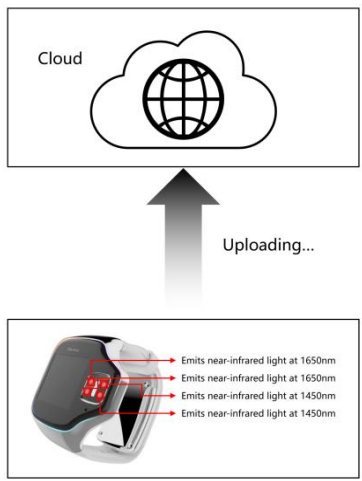

(a)

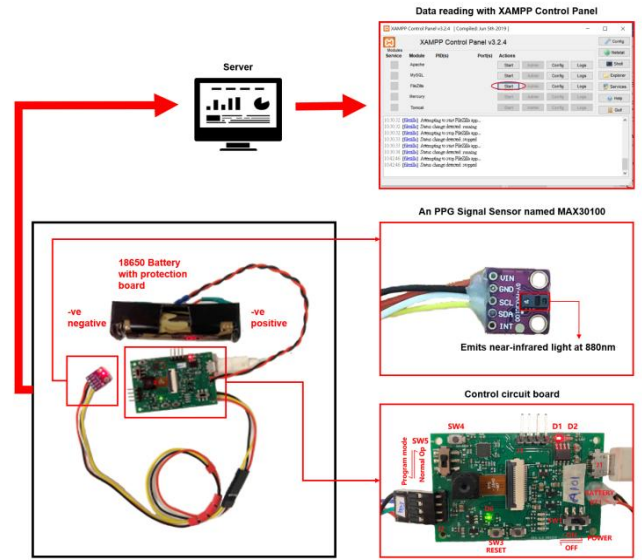

(b)

*Figure 5. a) The first acquisition device emitting the lights with their wavelengths being equal to 1450nm and 1650nm. b) The second acquisition device emitting the light with its wavelength being equal to 880nm.*

In order to obtain the reference blood glucose values and the reference blood pressure values, the Bene Check blood glucose meter and the Yuwell sphygmomanometer are employed for the acquisitions, respectively. It is worth noting that both the devices have the corresponding FDA certifications. In this paper, three measurements are taken every time and every measurement only takes one reference blood glucose value as well as one systolic blood pressure (SBP) value and one diastolic blood pressure (DBP) value. Then, the average of these three reference blood glucose values as well as the average of these three SBP values and the average of these three DBP values are employed as the corresponding final reference values.

2.3.2. Definition of the training set and the test set

To perform the training, all the PPGs acquired from an individual subject are mixed together. Then, the individual dataset is randomly divided into two non-overlapped subsets. They are the training set and the test set. In particular, approximately 75% of the individual dataset is defined as the training set and the rest approximately 25% of the individual dataset is defined as the test set. Figure 6 shows the exact total numbers of the PPGs in both the training set and the test set based on an individual subject.

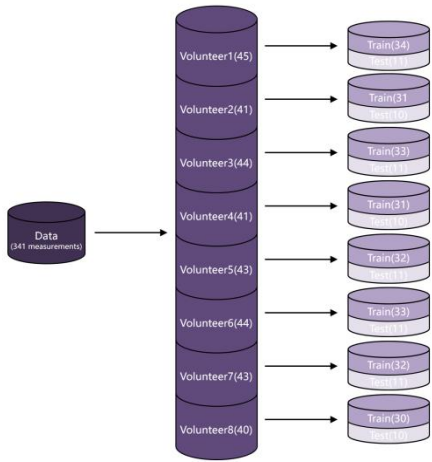

*Figure 6. The exact total numbers of the PPGs in both the training set and the test set based on the various individual subjects.*

2.3.3. Denoising

Since the heart rates of different subjects are different, the frequency bands of different PPGs acquired from different subjects are different. Hence, an adaptive based denoising method is required to address this problem. Among them, the singular spectrum analysis (SSA) based denoising method is the most common adaptive denoising method. However, the conventional SSA based denoising method is to discard the whole SSA component. In this case, some signal information is lost. To address this issue, an SSA based bit plane method is employed for denoising the PPGs. First, the SSA is applied to all the PPGs in both the training set and the test set. Here, the SSA decomposes each PPG into 32 components. Second, for each value in each SSA component, it is represented using a finite number of bits. In this paper, the wordlength of each value is 13 bits. Third, the noise level of each SSA component is estimated. Fourth, the total number of bits corresponding to the noise contaminated to each SSA component is set as the logarithm (based two) of its estimated noise level. Fifth, these noise bits are set to zero. Sixth, each value in each SSA component is reconstructed using the retained bits. Seventh, each PPG is reconstructed by summing up all the processed SSA components together. Eighth, the SSA is performed again on each reconstructed PPG. Ninth, only the first SSA component is retained and it is taken as the denoised PPG. Figure 7 shows a realization of the raw PPG and the corresponding denoised PPG. It can be seen from Figure 7 that the noise is significantly suppressed after performing the denoising.

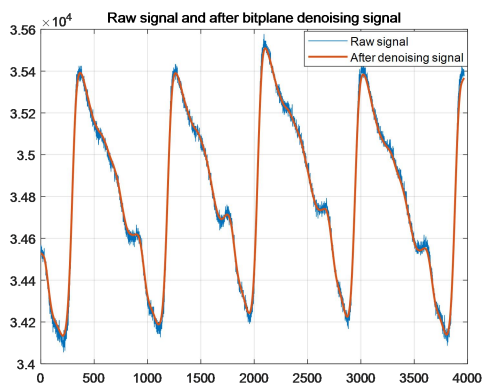

*Figure 7. A realization of the raw PPG and the corresponding denoised PPG.*

2.3.4. Feature extraction

103 features are extracted in each denoised PPG in both the training set and the test set. These 103 features are classified into five categories.

Category 1 (Features related to the meal time): First, each meal time and each data acquisition time are recorded in each measurement. Here, all the times are defined as the total number of hours after the midnight of that day. Then, the difference between the data acquisition time and the previous meal time is computed. This difference is taken as the feature. Therefore, there is only one feature for each feature vector in this category.

Category 2 (Features related to the blood pressure): Both the SBP value and the DBP value are employed as the features. Therefore, there are two features for this category.

Category 3 (Features related to the absorption of the infrared light by the blood glucose): There are two LEDs emitting the light with the wavelength being equal to 1450nm and another two LEDs emitting the light with the wavelength being equal to 1650nm. Therefore, there are four PPGs acquired at each time. Since the AC component and the DC component of the acquired PPGs are related to the absorption of the infrared light by the blood glucose, the mean and the variance of each PPG are computed and they are used as the features. As a result, there are eight features for each feature vector in this category.

Category 4 (Features related to the HRV): There is a LED emitting the light with the wavelength being equal to 880nm. Here, there are four finger cots covering the LED. Hence, there are also four PPGs acquired at each measurement. For each PPG, there are 13 features related to the HRV with six features extracted from the time domain and seven features extracted from the frequency domain. For those six features extracted from the time domain, they are the average and the standard deviation (SD) of the PP intervals, the root mean squares and the SD of the differences between two consecutive PP intervals as well as the total number and the percentage of the total number of the differences between two consecutive PP intervals with the differences being greater than 50ms. For those seven features extracted from the frequency domain, they are the total power of the PPG in the whole frequency spectrum, that in the frequency band between 0Hz and 0.04Hz (VLF), that in the frequency band between 0.04Hz and 0.15Hz (LF) and that in the frequency band between 0.15Hz and 0.4Hz (HF), as well as the ratio of the LF to the HF, the normalized LF and the normalized HF. Overall, there are 52 features for each feature vector in this category.

Category 5 (Features related to the HR): The same four PPGs defined in Category 4 are employed for extracting the HR features. First, the HRs are estimated based on the PP intervals of each PPG. Then, the mean, the median, the mode, the variance, the SD, the range, the interquartile range, the skewness, the kurtosis and the median absolute deviation of the HRs are computed. They are used as the features. Here, there are ten features for each PPG. As there are four PPGs for each measurement, there are 40 features for each feature vector in this category.

2.3.5. Smoothing the feature values or the reference blood glucose values

To smooth the feature values or the reference blood glucose values, first they are normalized. In particular, let $\hat{\mathbf{y}}_i$ be the vector containing the $i^{th}$ feature extracted from all the PPGs in the training set. Then, $\hat{\mathbf{y}}_i$ is normalized to the unit energy vector. Let $g_i = \frac{1}{\sqrt{\hat{\mathbf{y}}_i^T \hat{\mathbf{y}}_i}}$ be the normalized gain. Let $\mathbf{y}_i$ be the normalized vector. That is, $\mathbf{y}_i = g_i \hat{\mathbf{y}}_i$. Likewise, for the $i^{th}$ feature extracted from a PPG in the test set, the feature value is multiplied to $g_i$ to obtain the normalized feature value.

Second, a polynomial fitting approach is employed for smoothing the features or the reference blood glucose values in the training set. It is worth noting that a large value of the polynomial order does not achieve the smoothing purpose. On the other hand, a small value of the polynomial order results to the over smoothing. Hence, this paper sets the order of the polynomial equal to 3. Moreover, three different formulations are employed for finding the coefficients of the polynomial. They are based on the $L_2$ norm optimization method, the $L_1$ norm optimization method and the $L_\infty$ norm optimization method. Let $N$ be the total number of the measurements in the training set. For each normalized feature, the normalized feature values and the reference blood glucose values are first sorted according to the ascending order of the reference blood glucose values. Let $(x_i, y_i)$ be the $i^{th}$ sorted pair of the normalized feature value and the corresponding reference blood glucose value. Let $a_i$ be the coefficients of the polynomial.

2.3.5.1. Smoothing the reference blood glucose values via the $L_2$ norm optimization method

Let

$$X = \begin{bmatrix} 1 & x_1 & \cdots & x_1^3 \\ 1 & x_2 & \cdots & x_2^3 \\ \vdots & \vdots & \ddots & \vdots \\ 1 & x_N & \cdots & x_N^3 \end{bmatrix}, \quad (1a)$$

$$a = \begin{bmatrix} a_0 \\ \vdots \\ a_3 \end{bmatrix} \quad (1b)$$

and

$$y = \begin{bmatrix} y_1 \\ \vdots \\ y_N \end{bmatrix}. \quad (1c)$$

Let $J_2(a)$ be the objective function of the optimization problem. Here, it is defined as the energy of the error between the estimated blood glucose values and the actual blood glucose values. In particular, the optimization problem is formulated as follow:

$$\min_a J_2(a) = \sum_{i=1}^{N} (a_0 + a_1 x_i + \ldots + a_3 x_i^3 - y_i)^2. \quad (1d)$$

This is equivalent to

$$\min_a J_2(a) = \|Xa - y\|^2. \quad (1e)$$

By computing its gradient vector with respect to $a$ and setting its gradient vector to the zero vector, we have:

$$a = (X^T X)^{-1} X^T y. \quad (1f)$$

Once $a$ is found, $y$ is replaced by $Xa$. This is the vector of the smoothed reference blood glucose values. The above procedures are repeated for each feature and the mean of these vectors of the smoothed reference blood glucose values are taken as the final vector of the smoothed reference blood glucose values.

2.3.5.2. Smoothing the reference blood glucose values via the $L_1$ norm optimization method

Let $J_1(a)$ be the objective function of the optimization problem. Here, it is defined as the sum of the absolute error between the estimated blood glucose values and the actual blood glucose values. In particular, the optimization problem is defined as follow:

$$\min_a J_1(a) = \sum_{i=1}^{N} |a_0 + a_1 x_i + \ldots + a_3 x_i^3 - y_i|. \quad (2a)$$

This is equivalent to

$$\min_a J_1(a) = \|Xa - y\|_1. \quad (2b)$$

Let $z$ be a dummy vector and let $\iota$ be a vector with all its elements being equal to one. The above optimization problem is equivalent to the following optimization problem:

$$\min_{(a,z)} \iota^T z, \quad (2c)$$

subject to $Xa - y \leq z$,

and $-Xa + y \leq z$.

Let $I_N$ be the $N \times N$ identity matrix. Let $\tilde{a} = \begin{bmatrix} a \\ z \end{bmatrix}$, $f = \begin{bmatrix} 0 \\ \iota \end{bmatrix}$, $\tilde{X} = \begin{bmatrix} X & -I_N \\ -X & -I_N \end{bmatrix}$ and $\tilde{y} = \begin{bmatrix} y \\ -y \end{bmatrix}$. The above optimization problem becomes the following standard linear programming problem:

$$\min_{\tilde{a}} f^T \tilde{a}, \quad (2d)$$

subject to $\tilde{X}\tilde{a} \leq \tilde{y}$.

The solution of this standard linear programming problem can be found via the simplex method. By extracting the first four elements in $\tilde{a}$ and multiplying $X$ in front of this extracted vector, the vector of the smoothed reference blood glucose values is obtained. Likewise, $y$ is replaced by this smoothed reference blood glucose values and the above procedures are repeated for each feature. Finally, the mean of these vectors of the smoothed reference blood glucose values are taken as the final vector of the smoothed reference blood glucose values.

2.3.5.3. Smoothing the reference blood glucose values via the $L_\infty$ norm optimization method

Let $J_\infty(a)$ be the objective function of the optimization problem. Here, it is defined as the maximum absolute error between the estimated blood glucose values and the actual blood glucose values. In particular, the optimization problem is defined as follow:

$$\min_{a} J_{\infty}(a) = \max_{i} |a_0 + a_1 x_i + \ldots + a_3 x_i^3 - y_i|. \quad (3a)$$

This is equivalent to

$$\min_{a} J_{\infty}(a) = \|Xa - y\|_{\infty}. \quad (3b)$$

Let $\varepsilon$ be a dummy scalar. The above optimization problem is equivalent to the following optimization problem:

$$\min_{(a,\varepsilon)} \varepsilon, \quad (3c)$$

subject to $Xa - y \leq \varepsilon \iota$,

and $-Xa + y \leq \varepsilon \iota$.

Let $\hat{a} = \begin{bmatrix} a \\ \varepsilon \end{bmatrix}$, $\hat{f} = \begin{bmatrix} 0 \\ 1 \end{bmatrix}$ and $\hat{X} = \begin{bmatrix} X & -\iota \\ -X & -\iota \end{bmatrix}$. The above optimization problem becomes the following standard linear programming problem:

$$\min_{\hat{a}} \hat{f}^T \hat{a}, \quad (3d)$$

subject to $\hat{X}\hat{a} \leq \tilde{y}$.

Likewise, the solution of this standard linear programming problem can be found via the simplex method. Then, by extracting the first four elements in $\hat{a}$ and multiplying $X$ in front of this extracted vector, the vector of the smoothed reference blood glucose values is obtained. Next, $y$ is replaced by this smoothed reference blood glucose values and the above procedures are repeated for each feature. Finally, the mean of these vectors of the smoothed reference blood glucose values are taken as the final vector of the smoothed reference blood glucose values.

2.3.5.4. Smoothing the feature values

To smooth the feature values, the above three methods are repeated by swapping the vector of the feature values and the vector of the reference blood glucose values. However, instead of iterating the above procedures for each feature in the above algorithms, now it is only require to perform one iteration in the above algorithms because there is only one vector of the reference blood glucose values.

2.3.6. Feature selection

In this paper, the random forest (RF) is employed to rank the importance of the features. The features with the highest 25 importance are selected. Hence, the new training set and the new test set with 25 dimensional feature vectors are formed. It is worth noting that the individual modelling approach is employed. Therefore, different features are selected for different subjects.

2.3.7. Regression models

To develop the regression models, three models are employed. They are the support vector regression (SVR) model, the Gaussian regression model and the RF regression model. The parameters in these three models are found using the same training set.

2.3.8. Fusion of the various models

To yield a more accurate blood glucose estimation, the features or the reference blood glucose values smoothed by the polynomials with their coefficients found via the $L_1$ norm optimization method, the $L_2$ norm optimization method and the $L\infty$ norm optimization method are fused together. In particular, $\frac{2}{3}$ of the data in the previous training set is redefined as the new training set and the rest $\frac{1}{3}$ of the data in the previous training set is defined as the validation set. Therefore, 50% of the overall data is used to form the new training set, 25% of the overall data is used to form the validation set and the rest 25% of the overall data is used to form the test set. First, a RF based regression model is built using the feature vectors and the reference blood glucose values in the new training set with their values smoothed by the polynomial found via the $L_1$ norm optimization method. Then, the built model is applied to the feature vectors in the validation set for performing the blood glucose estimation. Next, the absolute differences between the estimated blood glucose values and the reference blood glucose values in the validation set are computed. After that, these absolute difference values are sorted according to the ascending order of the reference blood glucose values in the validation set. Let $e_1$ be the vector of these sorted absolute difference values. Likewise, the above procedures are repeated for the features or the reference blood glucose values smoothed by the polynomials found via the $L_2$ norm optimization method and the $L_\infty$ norm optimization method. Let $e_2$ and $e_\infty$ be the vectors of these sorted absolute difference values, respectively. Now, for each reference blood glucose value in the validation set, there are three absolute difference values based on the polynomials found by the $L_1$ norm optimization method, the $L_2$ norm optimization method and the $L_\infty$ norm optimization method. The optimization method corresponding to the minimum value among these three values is selected. It is found that the selected optimization methods can be partitioned into several regions according to the sorted reference blood glucose values. To determine the regions of the sorted reference blood glucose values and the corresponding selected optimization method, the accumulate probability approach is proposed. That is, the accumulate probability of each selected optimization method is computed. Let $\varepsilon$ be a threshold value. In this paper, $\varepsilon$ is set at 0.6. If the accumulate probability of any selected optimization method at a point is greater than $\varepsilon$, then the accumulate probabilities of these three selected optimization methods at that point are reset to zero and a range of the sorted blood glucose values are defined as that with the corresponding boundaries points being the previous reset point and this reset point. Hence, after performing the above procedures for all the reference blood glucose values in the validation set, the regions of the sorted reference blood glucose values and the corresponding optimization methods in these regions are determined. Finally, after applying the RF models formulated using these three optimization methods to a feature vector in the test set, there are three estimated blood glucose values. Then, the mean of these three estimated blood glucose values is computed. Next, according to the model developed in the validation set, the corresponding blood glucose value is selected as the final estimated blood glucose value.

## 3. Computer numerical simulation results

### 3.1. Results on various smoothing methods

As discussed in Section 2.3.5, the smoothing operations can be performed in the feature space or in the reference blood glucose space. This section determines the spaces whether the smoothing operations are performed or not via conducting the computer numerical simulations. Here, there are 10 different smoothing operations. They are only smoothing the feature values via the $L_1$ norm optimization method (denoted as O_f1), only smoothing the feature values via the $L_2$ norm optimization method (denoted as O_f2), only smoothing the feature values via the $L\infty$ norm optimization method (denoted as O_f∞), only smoothing the reference blood glucose values via the $L_1$ norm optimization method (denoted as B1_O), only smoothing the reference blood glucose values via the $L_2$ norm optimization method (denoted as B2_O), only smoothing the reference blood glucose values via the $L\infty$ norm optimization method (denoted as B∞_O), smoothing both the feature values and the reference blood glucose values via the $L_1$ norm optimization method (denoted as B1_f1), smoothing both the feature values and the reference blood glucose values via the $L_2$ norm optimization method (denoted as B2_f2), smoothing both the feature values and the reference blood glucose values via the $L\infty$ norm optimization method (denoted as B∞_f∞), and neither smoothing the feature values nor the reference blood glucose values (denoted as O_O).

In this paper, three different criteria are employed for evaluating the blood glucose estimation performances. They are the value of R, the mean absolute error (MAE) and the root mean squares error (RMSE). Here, the SD of the estimated blood glucose values is also presented. Table 1, Table 2 and Table 3 show the obtained performances yielded by various smoothing methods via the RF regression model, the SVR model and the Gaussian regression model, respectively. It can be seen from these tables that the RF regression model yields the best results in terms of all the above three criteria compared to the SVR model and the Gaussian regression model. Moreover, for the RF regression model, it can be seen from Table 1 that the O_f∞ method yields the best estimation result in terms of all the above three criteria. On the other hand, for both the SVR model and the Gaussian regression model, it can be seen from Table 2 and Table 3 that different methods would yield the best estimation results for different criteria. Hence, fusing different methods together would yield the better results.

**Table 1.** The values of R, MAE and RMSE yielded by the various smoothing methods via the RF regression model.

| Methods | R | MAE±SD (mmol/L) | RMSE (mmol/L) |
|---|---|---|---|
| O_f1 | 0.8959 | 0.9581±1.4041 | 1.1610 |
| O_f2 | 0.8743 | 0.8889±1.5229 | 1.1509 |
| O_f∞ | **0.9103** | **0.7991±1.8746** | **1.1319** |
| B1_O | 0.7834 | 0.9669±1.8230 | 1.1692 |
| B2_O | 0.7669 | 1.0080±1.8225 | 1.1673 |
| B∞_O | 0.7699 | 1.0160±1.9302 | 1.1772 |
| B1_f1 | 0.7895 | 1.0923±1.8227 | 1.1709 |
| B2_f2 | 0.7966 | 1.0285±1.8150 | 1.1685 |
| B∞_f∞ | 0.7576 | 1.0158±1.9332 | 1.1769 |
| O_O | 0.7545 | 1.1782±1.9416 | 1.1908 |

**Table 2.** The values of R, MAE and RMSE yielded by the various smoothing methods via the SVR model.

| Methods | R | MAE±SD (mmol/L) | RMSE (mmol/L) |
|---|---|---|---|
| O_f1 | 0.7402 | **0.9443±1.9292** | 1.1955 |
| O_f2 | **0.7917** | 1.0752±1.9659 | 1.1891 |
| O_f∞ | 0.5361 | 1.2745±2.4195 | 2.3640 |
| B1_O | 0.7156 | 1.0558±1.8306 | 1.1879 |
| B2_O | 0.7113 | 1.0968±1.8245 | 1.1867 |
| B∞_O | 0.6777 | 1.1017±1.9304 | 1.1953 |
| B1_f1 | 0.6622 | 1.1144±1.8095 | 1.1940 |
| B2_f2 | 0.7279 | 0.9757±1.8335 | 1.1848 |
| B∞_f∞ | 0.6197 | 1.1973±1.8862 | 1.2986 |
| O_O | 0.7411 | 0.9978±1.8770 | **1.1839** |

**Table 3.** The values of R, MAE and RMSE yielded by the various smoothing methods via the Gaussian regression model.

| Methods | R | MAE±SD (mmol/L) | RMSE (mmol/L) |
|---|---|---|---|
| O_f1 | 0.7146 | **0.9169±1.9050** | 1.1907 |
| O_f2 | 0.6263 | 1.1264±2.0025 | 1.2011 |
| O_f∞ | 0.6813 | 1.1086±1.9091 | 1.1965 |
| B1_O | 0.6998 | 1.0938±1.8078 | 1.1897 |
| B2_O | **0.7270** | 1.0050±1.8243 | **1.1850** |
| B∞_O | 0.6646 | 1.1179±1.9403 | 1.1983 |
| B1_f1 | 0.6822 | 1.1007±1.8372 | 1.1911 |
| B2_f2 | 0.6831 | 1.1013±1.8404 | 1.1900 |
| B∞_f∞ | 0.6523 | 1.1446±1.9513 | 1.1993 |
| O_O | 0.6064 | 1.2086±2.1098 | 1.2252 |

In order to demonstrate the effectiveness of fusing the various methods and the effectiveness of using the individual modelling approach, Table 4 shows the values of R, MAE, RMSE and MARD yielded by the fusion method, the O_f1 method, the O_f2 method, the O_f∞ method and the O_O method using the PPGs based on the individual subjects as well as the O_O method using the PPGs based on all the subjects. Figure 8 shows the order pairs of the estimated blood glucose values and the actual blood glucose values falling in the various regions in the Clarke error grid yielded by the various methods. Table 5 shows the percentages of these order pairs falling in the various regions in the Clarke error grid [25] yielded by different methods. It can be seen from Figure 8 that the O_f1 method, the O_f2 method and the O_f∞ method yield the best estimation performance in the low blood glucose range, the normal blood glucose range and the high blood glucose range, respectively. On the other hand, since the fusion method can enjoy the advantages of the individual methods, it can be seen from Table 4 and Table 5 that the fusion method yields the overall best performance. This demonstrates the advantage of using the fusion method. Moreover, since the individual modelling approach can eliminate the effects of different subjects having different responses on the infrared light, it can be seen from Table 4 and Table 5 that the O_O method using the PPGs from the individuals yields the better performance than those using the PPGs from all the subjects. This demonstrates the advantage of using the individual modelling approach.

**Table 4.** The values of R, MAE, RMSE and MARD yielded by the various methods.

| Methods | R | MAE±SD (mmol/L) | RMSE (mmol/L) | MARD |
|---|---|---|---|---|
| Fusion (PPGs based on the individuals) | **0.9466** | **0.6715± 1.8887** | **1.1068** | **0.0930** |
| O_f1 (PPGs based on the individuals) | 0.8959 | 0.9581± 1.4041 | 1.1610 | 0.1259 |
| O_f2 (PPGs based on the individuals) | 0.8743 | 0.8889± 1.5229 | 1.1509 | 0.1223 |
| O_f∞ (PPGs based on the individuals) | 0.9103 | 0.7991± 1.8746 | 1.1319 | 0.1101 |
| O_O (PPGs based on the individuals) | 0.8515 | 0.9388± 1.4757 | 1.1631 | 0.1252 |
| O_O (PPGs based on all the subjects) | 0.8277 | 0.9869± 1.4028 | 1.1701 | 0.1471 |

**Table 5.** The percentages of the order pairs of the estimated blood glucose values and the actual blood glucose values falling in the various regions in the Clarke error grid yielded by the various methods.

| Methods | A | B | C | D | E |
|---|---|---|---|---|---|
| Fusion (PPGs based on the individuals) | **94.1176%** | 5.882% | 0% | 0% | 0% |
| O_f1 (PPGs based on the individuals) | 83.5294% | 16.4706% | 0% | 0% | 0% |
| O_f2 (PPGs based on the individuals) | 84.7059% | 15.2941% | 0% | 0% | 0% |
| O_f∞ (PPGs based on the individuals) | 82.352% | 17.6471% | 0% | 0% | 0% |
| O_O (PPGs based on the individuals) | 80% | 18.8235% | 0% | 0% | 0% |
| O_O (PPGs based on all the subjects) | 77.647% | 22.3529% | 0% | 0% | 0% |

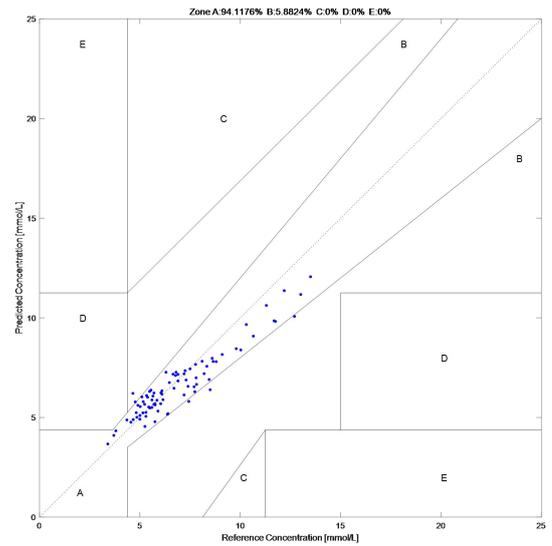
(a)

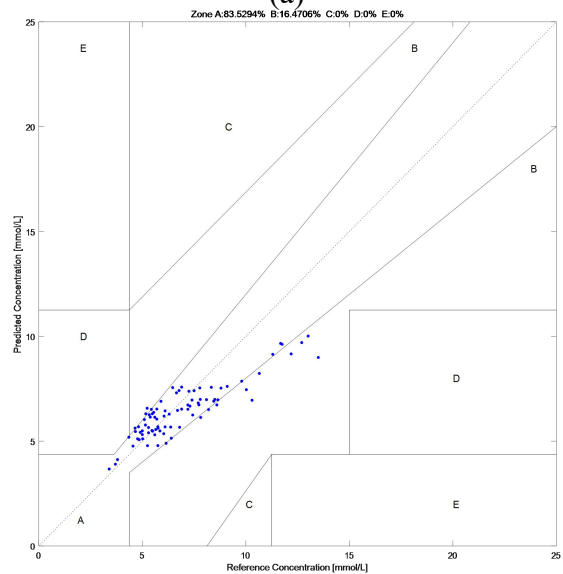
(b)

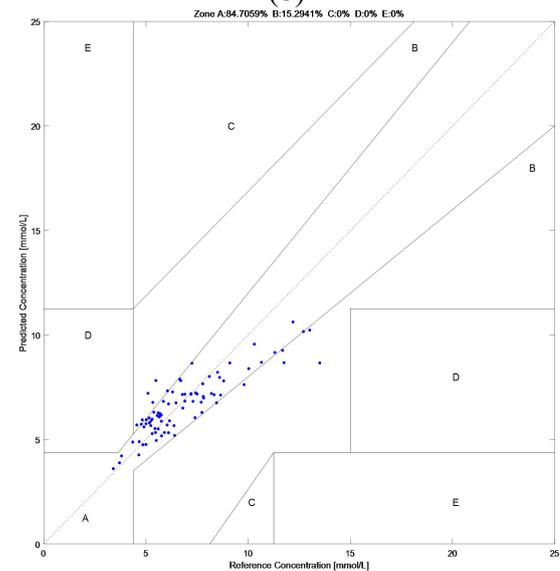
(c)

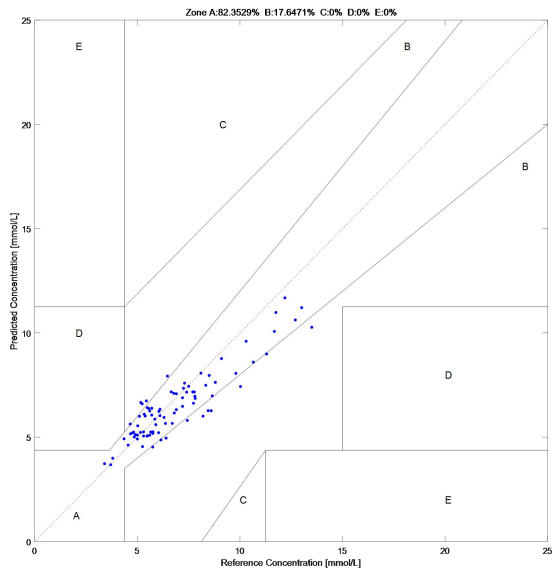
(d)

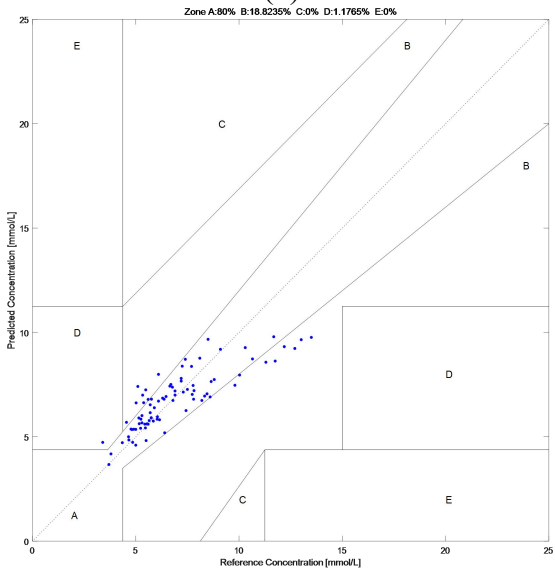
(e)

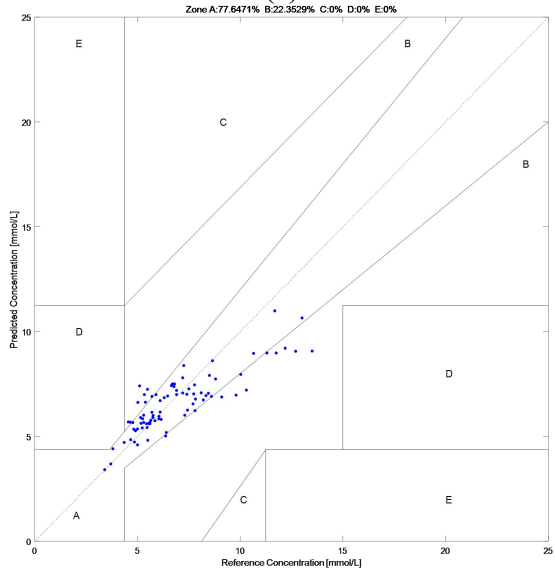
(f)

***Figure 8.*** *The order pairs of the estimated blood glucose values and the actual blood glucose values falling in the various regions in the Clarke error grid yielded by the various methods. (a) The fusion method using the PPGs based on the individuals. (b) The O_f1 method using the PPGs based on the individuals. (c) The O_f2 method using the PPGs based on the individuals. (d) The O_f∞ method using the PPGs based on the individuals. (e) The O_O method using the PPGs based on the individuals. (f) The O_O method using the PPGs based on all the subjects.*

3.2. Comparison to the existing methods

In order to have a fair comparison, the results yielded by all the methods are based on the same dataset established in this paper. Moreover, in order to verify the effectiveness of our proposed method, the results yielded by our proposed method are compared to those yielded by the current state of art method. In particular, since the SG filtering based method can effectively extract the underlying trend of the data and it is also based on the polynomial fitting approach, it is widely used for smoothing the features. Hence, this method is employed for performing the comparison. Here, the SG filtering with the following models including the RF model, the back propagation neural network (BPNN) model and the gradient boosting decision tree (GBDT) model is compared. It computes the total error within the window and the weights of the polynomial are obtained by minimizing the total error via the least squares approach. Here, the window length is set at 5 and the order of the polynomial is set at 3. Table 6 shows the obtained results. Besides, the existing work which is the closest to our proposed method [34] is also compared. Table 7 shows the obtained results. It can be seen from Table 6 that our proposed method outperforms the SG method for all these three models. Also, it can be seen from Table 7 that our proposed method is superior to the existing work [34]. In particular, our proposed method can achieve the value of R at 0.9466, the MAE at 0.6715, the RMSE at 1.1068, the MARD at 0.0930 and the percentage of the test data falling in the region A of the Clarke error grid at 94.1176%. Moreover, the total time required for performing both the training and the testing based on our proposed algorithm is low.

## 4. Conclusion

For the non-invasive blood glucose estimation, the LEDs emit the near infrared light and the near infrared light is received by the sensors for generating the PPGs. At the same time, the reference blood glucose values are taken. Then, the features are extracted from the PPGs. Next, the regression model is built based on the extracted features and the reference blood glucose values. Finally, for a given test data, the blood glucose value is estimated using the built model. However, due to the uncertainty of the acquisition device, the presence of the noise, the variations of the acquisition environments and different subjects having different responses of the infrared light to the blood glucose, the obtained features and the reference blood glucose values are highly unreliable. To address this issue, this paper proposes a polynomial fitting approach to smooth the obtained features or the reference blood glucose values to improve their reliability. In particular, the design of the weights in the polynomial is formulated as the various optimization problems with the objective functions of different optimization problems being the error function having different the norm criteria. Then, the results yielded by the various methods are fused together. Here, the training sets and the test sets of the regression models are defined based on the individual subjects. The computer numerical simulation results show that our proposed method yields 94.1176% of the data points falling in the region A of the Clarke error grid and the MARD at 0.0930.

Since the models are built by the individuals, the limitation of our proposed method is on the change of the physiological properties of the individuals such as the change of the medications taken by the individuals or the change of their meal habits. In this case, the correlation between the training set and the test set will be small. Like most of the machine learning algorithms, the estimation results will be poor. In future, the relationships among the individual models and the transfer from one individual model to another individual model will be investigated. Besides, the method will be developed for other biomedical signal processing applications such as the non-invasive blood lipid estimation.

**Table 6.** The values of R, MAE, RMSE and MARD, the percentages of the test data points falling in the region A and the region B of the Clarke error grid as well as the total time required for performing both the training and the testing yielded by our proposed method and the SG smoothing method with different models.

| Methods | R | MAE±SD (mmol/L) | RMSE (mmol/L) | MARD | Percentages of the test data points falling in the region A and the region B of the Clarke error grid | Total time required for performing both the training and the testing (s) |
|---|---|---|---|---|---|---|
| Fusion of the various optimization methods | **0.9466** | **0.6715±1.8887** | **1.1068** | 0.0930 | **94.1176% (region A) 5.8824% (region B)** | 0.3572s |
| SG with the RF model | 0.7326 | 0.9676±1.9917 | 1.1754 | 0.1284 | 84.6154% (region A) 15.3846% (region B) | 0.2794s |
| SG with the BPNN model | 0.4565 | 1.2093±2.4569 | 1.3728 | 0.1890 | 67.3077% (region A) 32.6923% (region B) | 0.6971s |
| SG with the GBDT model | 0.6213 | 1.0876±2.1365 | 1.1815 | 0.1432 | 82.6923% (region A) 17.3077% (region B) | **2.3045s** |

**Table 7.** The values of R, MAE, RMSE and MARD, the percentages of the test data points falling in the region A and the region B of the Clarke error grid as well as the total time required for performing both the training and the testing yielded by our proposed method and the existing work [34].

| Methods | R | MAE±SD (mmol/L) | RMSE (mmol/L) | MARD | Percentages of the test data points falling in the region A and the region B of the Clarke error grid | Total time required for performing both the training and the testing (s) |
|---|---|---|---|---|---|---|
| Fusion of the various optimization methods | **0.9466** | **0.6715±1.8887** | **1.1068** | 0.0930 | **94.1176% (region A) 5.8824% (region B)** | **0.3572s** |
| Existing work [34] | 0.8891 | 0.8761±1.9708 | 1.2617 | 0.1188 | 87.0588% (region A) 12.9412% (region B) | 4.6412s |

**Funding information**

This paper was supported partly by the National Nature Science Foundation of China (no. U1701266, no. 61671163, no. 62071128 and no. 61901123), the Team Project of the Education Ministry of the Guangdong Province (no. 2017KCXTD011), the Guangdong Higher Education Engineering Technology Research Center for Big Data on Manufacturing Knowledge Patent (no. 501130144) and the Hong Kong Innovation and Technology Commission, Enterprise Support Scheme (no. S/E/070/17).

**Conflict of interest**

There is no conflict of interest.

**Credit contribution statement**

Yiting Wei is responsible for developing the methodology, implementing the algorithm, acquiring the data and writing the draft of the paper.

Bingo Wing-Kuen Ling is responsible for attracting the funding, managing the project, developing the methodology, supervising the project and revising the paper.

Danni Chen, Yuheng Dai and Qing Liu are responsible for validatiing the results and acquiring the data.